\def\year{2022}\relax
\documentclass[letterpaper]{article} 
\usepackage{paralist}
\usepackage{amsmath}

\usepackage{aaai22}  
\usepackage{times}  
\usepackage{helvet} 
\usepackage{courier}  
\usepackage[hyphens]{url}  
\usepackage{graphicx} 
\usepackage{natbib}  
\usepackage{caption} 
\usepackage{xspace}
\usepackage{booktabs}
\usepackage{amsmath}
\usepackage{tabularx}

\usepackage[utf8]{inputenc}
\usepackage[T1]{fontenc}
\usepackage{hyphenat}
\usepackage{xspace}
\usepackage{amsmath}
\usepackage{amsfonts}
\usepackage{hyperref}
\usepackage{url}
\usepackage{booktabs}
\usepackage{multirow}
\usepackage{subfigure}
\usepackage{makecell}
\usepackage{caption}
\usepackage{minibox}
\usepackage{bbm}
\usepackage{graphicx}
\usepackage{balance}
\usepackage{mathtools}
\usepackage{color}
\usepackage{marvosym}
\usepackage{ifthen}
\usepackage{textcomp}
\usepackage{enumitem}
\usepackage{verbatim}
\usepackage{algorithm}
\usepackage{algorithmic}
\usepackage{numprint}
\usepackage{balance}
\usepackage{amsthm}

\newcommand{\chatoDisplayMode}[1]{#1}

\definecolor{MyRed}{rgb}{0.6,0.0,0.0} 
\definecolor{MyBlack}{rgb}{0.1,0.1,0.1} 
\newcommand{\inred}[1]{{\color{MyRed}\sf\textbf{\textsc{#1}}}}
\newcommand{\frameit}[2]{
  \begin{center}
  {\color{MyRed}
  \framebox[.9\columnwidth][l]{
    \begin{minipage}{.85\columnwidth}
    \inred{#1}: {\sf\color{MyBlack}#2}
    \end{minipage}
  }\\
  }
  \end{center}
}

\newcommand{\note}[2][]{\chatoDisplayMode{\def\@tmpsig{#1}\frameit{{\Pointinghand} Note}{#2\ifx \@tmpsig \@empty \else \mbox{ --\em #1}\fi}}}
\newcommand{\todo}[2][]{\chatoDisplayMode{\def\@tmpsig{#1}\frameit{{\Writinghand} To-do}{#2\ifx \@tmpsig \@empty \else \mbox{ --\em #1}\fi}}}

\newcommand{\abbrevStyle}[1]{#1}

\newcommand{\ie}{\abbrevStyle{i.e.}\xspace}
\newcommand{\eg}{\abbrevStyle{e.g.}\xspace}

\newcommand{\vs}{\abbrevStyle{vs.}\xspace}

\newcommand{\Secref}[1]{Sec.~\ref{#1}}

\newcommand{\Tabref}[1]{Table~\ref{#1}}
\newcommand{\Figref}[1]{Fig.~\ref{#1}}

\newcommand{\xhdr}[1]{\vspace{1.7mm}\noindent{{\bf #1.}}}

\newcommand{\xhdrNoPeriod}[1]{\vspace{1.7mm}\noindent{{\bf #1}}}

\newcommand{\textcite}[1]{\citeauthor{#1} \shortcite{#1}}

\newcommand{\hide}[1]{}

\makeatletter
\newcommand{\iffont}[2]{\ifthenelse{\equal{\f@family}{#1}}{#2}{}}
\makeatother

\iffont{ptm}{
  \usepackage{mathptmx}

  \DeclareSymbolFont{greek}{OML}{cmm}{m}{n}
  \DeclareMathSymbol{\alpha}{\mathalpha}{greek}{"0B}
  \DeclareMathSymbol{\beta}{\mathalpha}{greek}{"0C}
  \DeclareMathSymbol{\gamma}{\mathalpha}{greek}{"0D}
  \DeclareMathSymbol{\delta}{\mathalpha}{greek}{"0E}
  \DeclareMathSymbol{\epsilon}{\mathalpha}{greek}{"0F}
  \DeclareMathSymbol{\zeta}{\mathalpha}{greek}{"10}
  \DeclareMathSymbol{\eta}{\mathalpha}{greek}{"11}
  \DeclareMathSymbol{\theta}{\mathalpha}{greek}{"12}
  \DeclareMathSymbol{\iota}{\mathalpha}{greek}{"13}
  \DeclareMathSymbol{\kappa}{\mathalpha}{greek}{"14}
  \DeclareMathSymbol{\lambda}{\mathalpha}{greek}{"15}
  \DeclareMathSymbol{\mu}{\mathalpha}{greek}{"16}
  \DeclareMathSymbol{\nu}{\mathalpha}{greek}{"17}
  \DeclareMathSymbol{\xi}{\mathalpha}{greek}{"18}
  \DeclareMathSymbol{\pi}{\mathalpha}{greek}{"19}
  \DeclareMathSymbol{\rho}{\mathalpha}{greek}{"1A}
  \DeclareMathSymbol{\sigma}{\mathalpha}{greek}{"1B}
  \DeclareMathSymbol{\tau}{\mathalpha}{greek}{"1C}
  \DeclareMathSymbol{\upsilon}{\mathalpha}{greek}{"1D}
  \DeclareMathSymbol{\phi}{\mathalpha}{greek}{"1E}
  \DeclareMathSymbol{\chi}{\mathalpha}{greek}{"1F}
  \DeclareMathSymbol{\psi}{\mathalpha}{greek}{"20}
  \DeclareMathSymbol{\omega}{\mathalpha}{greek}{"21}
  \DeclareMathSymbol{\varepsilon}{\mathalpha}{greek}{"22}
  \DeclareMathSymbol{\vartheta}{\mathalpha}{greek}{"23}
  \DeclareMathSymbol{\varpi}{\mathalpha}{greek}{"24}
  \DeclareMathSymbol{\varrho}{\mathalpha}{greek}{"25}
  \DeclareMathSymbol{\varsigma}{\mathalpha}{greek}{"26}
  \DeclareMathSymbol{\varphi}{\mathalpha}{greek}{"27}
  \DeclareSymbolFont{otone}{OT1}{cmr}{m}{n}
  \DeclareMathSymbol{\Gamma}{\mathalpha}{otone}{0}
  \DeclareMathSymbol{\Delta}{\mathalpha}{otone}{1}
  \DeclareMathSymbol{\Theta}{\mathalpha}{otone}{2}
  \DeclareMathSymbol{\Lambda}{\mathalpha}{otone}{3}
  \DeclareMathSymbol{\Xi}{\mathalpha}{otone}{4}
  \DeclareMathSymbol{\Pi}{\mathalpha}{otone}{5}
  \DeclareMathSymbol{\Sigma}{\mathalpha}{otone}{6}
  \DeclareMathSymbol{\Upsilon}{\mathalpha}{otone}{7}
  \DeclareMathSymbol{\Phi}{\mathalpha}{otone}{8}
  \DeclareMathSymbol{\Psi}{\mathalpha}{otone}{9}
  \DeclareMathSymbol{\Omega}{\mathalpha}{otone}{10}
  \DeclareSymbolFont{syms}{OML}{cmm}{m}{it}
  \DeclareMathSymbol{\partial}{\mathord}{syms}{"40}
  \DeclareMathAlphabet{\mathbold}{OML}{cmm}{b}{it}
  \DeclareSymbolFont{largesymbols}{OMX}{cmex}{m}{n}

}
\usepackage{hyphenat}
\usepackage{mathptmx}

\newif\ifcomment
\commenttrue
\ifcomment
\newcommand{\sz}[1]{{\bf \textcolor{blue}{SZ: #1}}}
\newcommand{\oana}[1]{{\color{red}#1 --\textbf{Oana}}}

\else
\newcommand{\sz}[1]{}
\newcommand{\oana}[1]{}

\fi

\title{Can online attention signals help fact-checkers  fact-check?}

\author
{Manoel Horta Ribeiro,$^1$ 
Savvas Zannettou,$^2$
Oana Goga,$^3$
Fabrício Benevenuto,$^4$
Robert West$^1$\\
}
\affiliations{
manoel.hortaribeiro@epfl.ch, 
szannett@mpi-inf.mpg.de,
oana.goga@cnrs.fr,
fabricio@dcc.ufmg.br,
robert.west@epfl.ch \\
$^1$EPFL, 
$^2$TU Delft, 
$^3$Univ. Grenoble Alpes, 
$^4$UFMG
}

\begin{document}

\maketitle

\begin{abstract}
Recent research suggests that not all fact-checking efforts are equal: \emph{when} and \emph{what} is fact-checked plays a pivotal role in effectively correcting misconceptions.
In that context, signals capturing how much attention specific topics receive on the Internet have the potential to study (and possibly support) fact-checking efforts.
This paper proposes a framework to study fact-checking with online attention signals. The framework consists of: 
1)~extracting claims from fact-checking efforts; 
2)~linking such claims with knowledge graph entities; and
3)~estimating the online attention these entities receive.
We use this framework to conduct a preliminary study of a dataset of 879 COVID\hyp{}19\hyp{}related fact-checks done in 2020 by 81 international organizations.
Our findings suggest that there is often a disconnect between online attention and fact-checking efforts.
For example, in around 40\% of countries that fact-checked ten or more claims, half or more than half of the ten most popular claims were not fact-checked.
Our analysis also shows that claims are first fact-checked after receiving, on average, 35\% of the total online attention they would eventually receive in 2020.
Yet, there is a considerable variation among claims:
some were fact-checked before receiving a surge of misinformation-induced online attention;
others are fact-checked much later.
Overall, our work suggests that the incorporation of online attention signals may help organizations assess their fact-checking efforts and choose what and when to fact-check claims or stories.
Also, in the context of international collaboration, where claims are fact-checked multiple times across different countries, online attention could help organizations keep track of which claims are ``migrating'' between countries.
\end{abstract}

\section{Introduction}

Recent events such as the 2016 U.S. elections and the COVID\hyp{}19 pandemic have placed fact-checking into the spotlight~\cite{graves_deciding_2016, krause2020fact}.
All across the globe, dozens of organizations dedicate themselves to verifying the accuracy of claims and stories circulating through our information ecosystem~\cite{poynter_br_2017, poynter_sk_2017}.
Ideally, such efforts would decrease the negative impact that false information has on people's lives as well as increase political accountability and democratic outcomes~\cite{nyhan2015effect}.

Empirical research suggests that fact-checking efforts are effective in correcting misconceptions~\cite{walter_fact-checking_2020}, but that their impact is transient and non-cumulative~\cite{nyhan_why_2021}.
Given this reality, researchers and fact-checking organizations have been exploring various strategies to increase the efficacy of fact-checking. 
These range from studying how and when fact-checking should be done~\cite{brashier_timing_2021, pennycook_shifting_2021} to creating institutions that can promote collaborations across countries and organizations~\cite{ifcn}. 

Signals that capture how much attention specific claims or stories receive on the Internet (where many read their news~\cite{shearer2021more}) can improve the efficacy of fact-checking~\cite{lazer_studying_2020}.
With these signals, organizations could fact-check specific claims or stories only after they had received enough attention (after all, fact-checking unknown stories may only bring them visibility~\cite{donovan2021stop}). 
Further, they could evaluate their fact-checking efforts, calculating, for instance,  how well they covered popular false stories through time.

\xhdr{Present work}
The promises of online attention data motivated us to explore the usage of Google Trends to characterize fact-checking efforts.
Google Trends is a freely available, ``always-on'' signal that captures the search volume received by specific queries in the world's largest search engine.
The signal is available on a country-level granularity (and sometimes district-level), and the usage of entity identifiers from Google's Knowledge Graph (GKG)~\cite{google-kg} makes it easy to replicate the same query across different countries/languages.

Using Google Trends as a proxy for online attention, we develop a framework where we: 
1)~semi-automatically extract claims from unstructured fact-checking data;
2)~link fact-checked claims with knowledge graph entities;
and 3)~estimate the misinformation-induced attention these claims received using Google Trends.
We leverage this framework to investigate two research questions:

\begin{itemize}
\item
\textit{RQ1:}
\textbf{Measuring the relevance of fact-checking.}
What are the differences in online attention between fact-checked and not fact-checked claims?
\item
\textit{RQ2:}
\textbf{Measuring the speed of fact-checking.} 
When in their attention life cycle are falsehoods fact-checked?
\end{itemize}

\xhdr{Application to COVID\hyp{}19\hyp{}related misinformation} 
Motivated by the importance of misinformation spread in the context of the  COVID\hyp{}19 pandemic, we apply the above framework to a dataset provided by the International Fact-checking Network (IFCN)~\cite{ifcn}.
The dataset contains metadata from 7,519 fact-checking efforts (\textit{fact-checks}) made between January and July 2020 by 95 organizations in 136 countries and regions.%
\footnote{Some fact-checks were attributed to geopolitical regions, \eg, the Middle East.} 
We use our methodology to extract 39 claims that were fact-checked 879 times in 72 countries and analyze 2,568 different Google Trends time series, each related to a different $\langle$country, claim$\rangle$ pair.

\xhdr{Summary of findings}
Fact-checked claims did not have the highest surge in search volume for many of the countries studied. 
For example, contrasting the fact-checked and not fact-checked claim that received the most attention in each country, the fact-checked claim was, on average, 44.4\% \emph{less} popular than the not fact-checked claim.
Countries with more fact-checks seem to do better: for countries where ten or more claims were fact-checked, the top most popular fact-checked claims consistently received more attention than the top most popular non-fact-checked claims.
Nevertheless, in around 40\% of countries where ten or more claims were fact-checked, half or more than half of the most popular claims were not fact-checked.
Overall, these findings suggest a disconnect between search volume and fact-checking behavior, which indicates that viral falsehoods were left unchecked in many countries.

We also find a high variance on \textit{when} claims get fact-checked in their attention life cycle. 
Claims were first fact-checked after receiving, on average, 35\% of the total attention they would eventually receive in 2020. 
Yet, while some claims are fact-checked when the falsehood is beginning to spread (\eg, coffee cures COVID\hyp{}19), others are fact-checked after the peak of attention, towards the time that the falsehood has almost faded out (\eg, a Harvard professor was arrested for creating COVID\hyp{}19).
These differences were significant even after controlling for the total attention received by claims and the dates when claims were first fact-checked in each country. 

\xhdr{Implications}
Our case study suggests that Google Trends may help to assess fact-checking efforts and to decide \textit{what} and \textit{when} to fact-check. 
More specifically, it could help ground fact-checking efforts in the information needs of individuals and allow news organizations to deliver fact-checks when they are most effective.
The signal can also be used to track the spread of specific falsehoods across the globe, helping in the collaborative efforts of international fact-checking networks like the IFCN. 

\section{Related Work}

Our work builds upon fast-growing literature studying online attention and fact-checking.

\xhdr{Online attention}
Much of the research studying online attention on the Internet has relied on Google Trends, a service that since 2006 offers access to aggregate Google search data. 
Google Trends has been used for purposes that range from ``nowcasting'' economic indicators~\cite{choi2012predicting} to quantifying the association between racism and black mortality in the United States~\cite{chae2015association}. 
\citet{jun2018ten} provide a thorough analysis of how the tool was used between 2006 and 2016. 
One problem with publicly available Google Trends is that it is rounded to the nearest integer and normalized per query (the highest value equals 100). 
Recently, \citet{gtab} has proposed Google Trends Anchor Bank (G-TAB), a methodology that allows researchers to overcome these limitations by using a set of predefined queries.

Another signal explored to study online attention is Wikipedia page views, made publicly available by the Wikimedia Foundation.
The signal has also been used for a variety of purposes, which range from studying changes in information-seeking behavior during the COVID\hyp{}19 pandemic~\cite{ribeiro2021sudden} to predicting election outcomes~\cite{smith2017using}.
The framework we propose in this paper could be easily extended to Wikipedia data, which spans dozens of language editions. 

A challenge associated with using online attention signals is that they are an imperfect proxy for the real world.
The story of Google Flu Trends~\cite{dugas2013influenza}, a flu tracking system that leveraged search data illustrates these limitations.
In 2013, the tool made headlines for drastically overestimating the presence of the flu in the 2012-2013 season~\cite{lazer2014parable}, highlighting the disconnect that can exist between online attention and the real world.

\xhdr{Fact-checking}
With the rise of fact-checking in the last decade, research has explored multiple facets of fact-checking efforts.
\citet{graves2016understanding} have analyzed the motivations behind fact-checking through a field experiment where they exposed journalists to messages encouraging the practice. 
Their findings suggest that political fact-checking is driven primarily by professional motives within journalism rather than by audience demand.
\citet{brashier_timing_2021} studied how timing influences the efficacy of fact-checking efforts in correcting misconceptions. 
In two controlled experiments, they observed that providing fact-checks after exposing participants to false news was more effective than providing them before exposure.
Several authors have analyzed the efficiency of fact-checks (\eg, see \citet{ecker2020effectiveness}). 
Overall, these studies find a significant effect in correcting misinformation~\cite{walter_fact-checking_2020}, but that their impact is transient and non-cumulative~\cite{nyhan_why_2021} and their success is conditioned on individuals' pre-existing beliefs~\cite{walter_fact-checking_2020,nyhan2010corrections}.

Parallel to the investigation of ``traditional'' fact-checking, researchers have also explored ways of partially or entirely automating the process~\cite{hassan2015quest,thorne2018automated}.
For example, \citet{rashkin2017truth} explore the linguistic characteristics of untrustworthy text and \citet{ciampaglia2015computational} leverage knowledge graphs to assess the veracity of statements.
Other work aims to develop automated methods that can be more explicitly incorporated into fact-checking practices.
In that direction, we highlight the work of \citet{hassan2017toward}, which develop a claim-spotting platform that aims to detect check-worthy factual claims in political discourses.

\xhdr{Relation to prior work}
Previous work has illustrated the usefulness and the limitations of using Google Trends as a signal to study and model real-world phenomena~\cite{choi2012predicting, lazer2014parable}.
Here, we take the opposite approach: we are interested in the online attention towards specific topics: the value being measured by the service.
We provide a framework to assess, in a given country: 1)~what claims were fact-checked, and 2)~when, relative to the claim attention life cycle, the fact-checking happened. 
We argue that this objective is orthogonal to the goals of the previously reviewed literature and may serve as a stepping stone for a variety of future work.
Google Trends data may be used in an observational fashion to test theories about what makes fact-checking effective and incorporate online attention into fact-checking.

\section{Methods}
\label{sec:methods}

\begin{table}[t]
\caption{Example of an entry in the dataset provided by IFCN. Here, a South African organization named "AfricaCheck" fact-checked the claim that Christiano Ronaldo would be turning his hotels into COVID\hyp{}19 hospitals. The claim was spotted in March 2020 and was considered false.}
\label{tab:fcex}
\begin{tabular}{p{3.5cm}|p{3.75cm}}
\toprule
\textbf{Date} & 15th of March of 2020  \\ \midrule
\textbf{Country} & South Africa \\ \midrule
\textbf{Organization} & AfricaCheck \\ \midrule
\textbf{What was \newline {fact-checked?}} & Footballer Ronaldo is turning his hotels into COVID-19 hospitals. \\ \midrule
\textbf{Who said/posted it?} & Facebook\\ \midrule
\textbf{Link to original} & (not available)\\ \midrule
\textbf{URL to article} & \url{https://bit.ly/3rKutq2}\\ \midrule
\textbf{Language of fact-check} & English\\ \midrule
\textbf{Final rating} & False\\ \midrule
\textbf{Explanation} & There is no evidence that Ronaldo's luxury hotels in Portugal are being turned into hospitals.\\ \bottomrule
\end{tabular}
\end{table}

This section describes our dataset and the developed framework to combine fact-checking data with Google Trends.

\subsection{Fact-checking Data} 
Our analysis uses a dataset with data on COVID-19-related fact-checking efforts provided by the International Fact-checking Network (IFCN)~\cite{ifcn}.
The dataset contains metadata from 7,519 fact-checking efforts (\textit{fact-checks}) made between January and July 2020 by 95 organizations in 136 countries and regions.%
The dataset includes COVID-19 false information that went viral across many countries (\eg, masks cause hypoxia), as well as very specific false claims that were disseminated on popular mainstream social networks like Facebook and Twitter (\eg, a specific image or video shared on Facebook/Twitter is fake).
For each fact-check, the dataset includes \emph{when} the fact-checker saw the specific claim, the fact-checking organization that fact-checked the claim, the countries where the fact-check was disseminated, and a description of the fact-check (all descriptions are in English, regardless of the fact-checking organization's country).

In \Tabref{tab:fcex}, we provide an example of an entry in this dataset. We note that this data is available online%
\footnote{\url{https://www.poynter.org/ifcn-covid-19-misinformation/}}
in a simplified form for a longer time frame (it is constantly updated). We make available all data used in this paper as well as code to reproduce our results%
\footnote{https://github.com/epfl-dlab/fact-checkers-fact-check}

\subsection{Google Trends} 
Google Trends allows the analysis of the popularity of search queries across time and countries in the world's largest search engine. 
Queries can be specified as plain text (\eg, ``hypoxia'') or as entity identifiers from Google's Knowledge Graph~(GKG)~\cite{google-kg} (\eg, hypoxia corresponds to the entity \texttt{/m/2F03gns/}). 
Importantly, by querying entity identifiers, we aggregate multiple queries related to the entity, and the method is robust to language variations of queries. 
For instance, even though hypoxia is written \textit{hipoxia} in Spanish and \textit{Hypoxie} in German, we can retrieve the popularity of the entity \texttt{/m/2F03gns/} on Google Trends without worrying about the translation of the word. 

One problem with Google Trends is that it yields normalized results.
As described in the Google Trends' Frequently Asked Questions section:\footnote{https://support.google.com/trends/answer/4365533?hl=en}
\textit{``Each data point is divided by the total searches of the geography and
time range it represents to compare relative popularity. [...] The
resulting numbers are then scaled on a range of 0 to 100 based on a
topic’s proportion to all searches on all topics.''}
To address this issue, which does not allow us to make comparisons across claims, we use G-TAB~\cite{gtab}, a method to calibrate Google Trends data that allows us to obtain queries on a universal scale and to minimize errors stemming from Google's rounding to the nearest integer. 
G\hyp{}TAB works by building country-specific ``anchor banks,'' sets of queries of varying popularity that are calibrated according to a common reference query. 
When posing a new query, G-TAB allows calibrating using the anchor bank and obtaining the search interest normalized by the reference query in the anchor bank.
Since all queries are expressed in the same unit (the reference query), we can meaningfully compare the search interest between as many queries as we want.
Note that we build a distinct ``anchor bank'' for each country. 
We discuss how to aggregate signals derived from G-TAB for measurements across countries later in the section.

\subsection{Combining Fact-checks with Google Trends Data}
\label{sec:fcgt}

\begin{figure}
\centering
\includegraphics{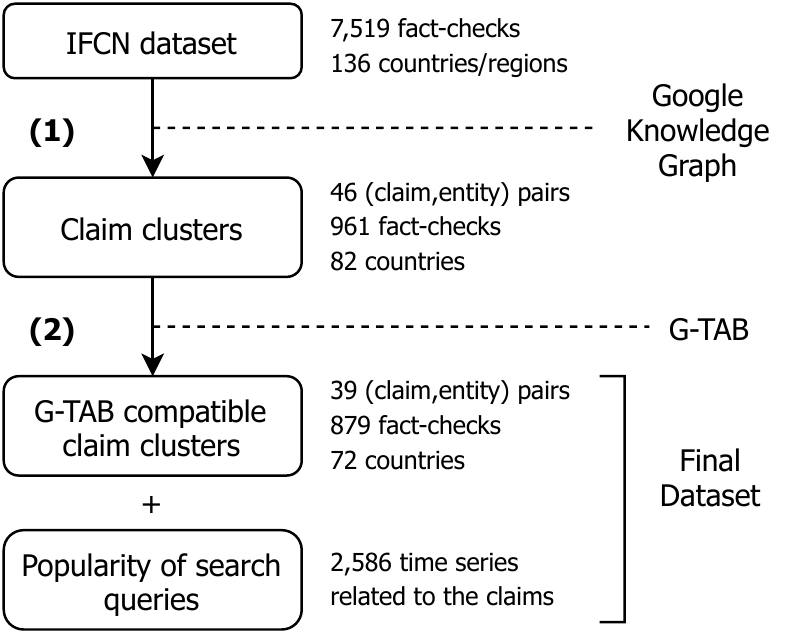}
\caption{Diagram illustrating the overall data collection pipeline. Steps (1) and (2) are explained in the main text.}
\label{fig:diag}
\end{figure}

To combine fact-checking data with Google Trends, we link fact-checks with knowledge graph entities and then query these knowledge graph entities across various countries using G-TAB. 
These steps, explained in detail below, are illustrated in \Figref{fig:diag}.

\begin{figure*}
\centering
\includegraphics[width=\linewidth]{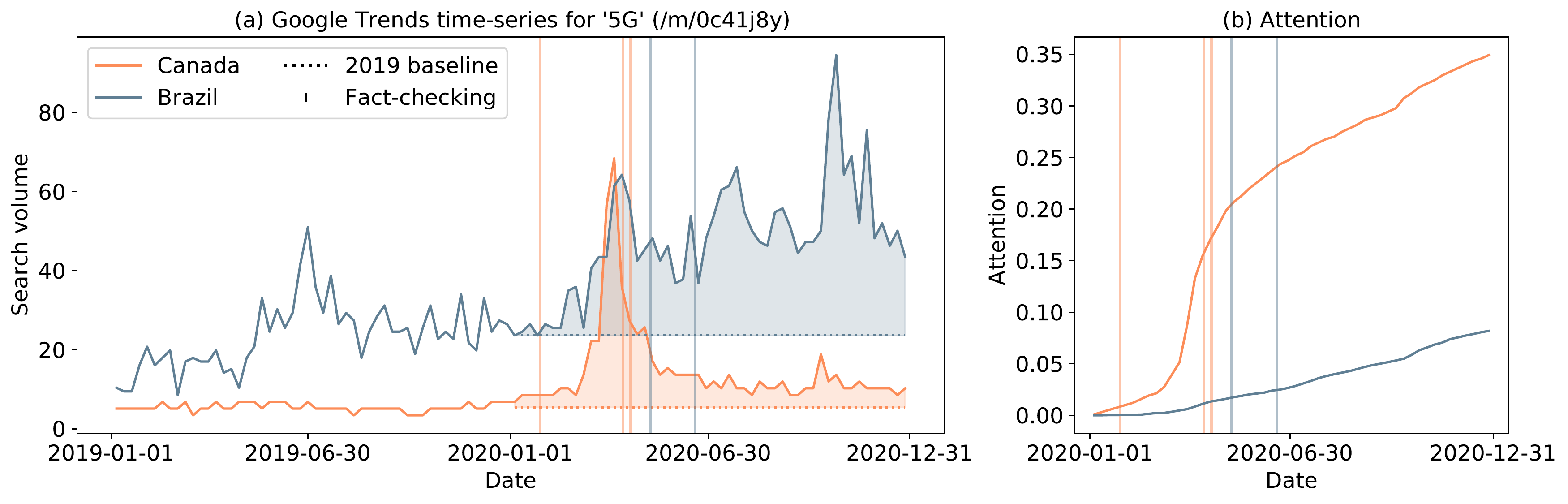}
\caption{In (a), we show the search volume time series for the entity ``5G'' for Canada and Brazil along with fact-checking efforts.
The dashed line represents the 2019 baseline, and the colored area represents the difference between the search volume and the baseline.
We depict fact-checks as vertical lines.
Here, the unit of the search volume is country-specific
In (b), we show the total attention associated with these entities.
Here, we have normalized the search volume across the two countries.
}
\label{fig:measures}
\end{figure*}

\xhdr{(1) Linking fact-checks with knowledge graph entities}
First, we create a Boolean bag-of-words vector representation for each fact-check using the textual descriptions provided in the IFCN dataset. We did not consider stop-words and did not distinguish between the upper and lower case when creating these representations.
We then cluster these fact-checks using DBSCAN~\cite{ester1996density}%
\footnote{We use the implementation from Scikit-learn~\cite{pedregosa2011scikit}}
 on the matrix of pairwise Jaccard distances between the bag-of-words vectors.
The DBSCAN algorithm mainly relies on the $\epsilon$ hyper-parameter, which defines the maximum distance between two samples in the same cluster.
To identify a good $\epsilon$ value, we experimented with various values between 0.3 and 0.7 (with 0.05 increments) and qualitatively assessed the quality of the clusters.
For our analysis, we use an $\epsilon$ value of 0.5, since it yielded the most cohesive clusters.
Second, we manually identify clusters corresponding to an entity in  Google Knowledge Graph (GKG).
For example, in one of the clusters obtained, we find fact-checks debunking the claim that ``alcohol cures coronavirus.''
We assign all fact-checks in this cluster with the entity ``alcoholic drink,'' (\texttt{m/012mj} in GKG).
To do this, we manually extract a representative word from the description of the claim (in this specific example, ``alcohol'') and then perform a search query using Google's Knowledge Graph Search API~\cite{kg-api}.
Then, we manually check the search results and select the entity closer to the representative word.
Clusters that are too noisy are not associated with any entity.
Two large clusters contained several claims related to either alleged cures (X cures COVID-19) or alleged predictions of the pandemic (Y predicted COVID-19). 
We break down these clusters before associating them with GKG entities.
Additionally, we searched for each of the GKG entities identified in the entire dataset to ensure that we did not miss any relevant fact-check; all fact-checks found this way are then associated with the clusters based on entities. 
Note that entities here are quite general, \eg ``alcoholic drink.'' 
We describe how we account for the misinformation-specific attention later in this section.

Overall, this process yielded 46 clusters, out of which we were able to associate 39 with a GKG entity.
In aggregate, the clusters contained 961 fact-checks from 83 organizations in 82 countries/regions. 
All clusters obtained contained only claims labeled by fact-checkers as false, misleading, partly false, or ``no evidence.''
We note that, while the linkage with GKG substantially reduced the data at hand, it is a crucial pre-processing step to subsequently combine the IFCN data with popularity time series extracted from Google Trends. 

\xhdr{(2) Measuring search interest with Google Trends} 
Next, we built G-TAB anchor banks for the countries associated with the fact-checks selected.
For this step, we discarded fact-checks associated with regions (\eg, Middle East) and countries where Google Trends was too noisy to build anchor banks (this happened for four countries: Malawi, South Korea, Timor, and Kyrgyzstan).
In total, we obtained anchor banks for 72 countries that have 879 fact-checks from 81 organizations. 
Then, we searched each of the entities associated with the 39 fact-checking clusters for each anchor bank.
Ideally, if all searches were successful, we would obtain 2,808 (39 claims $\times$ 72 countries). 
However, we only obtained 2,586 ($\sim$92\%) time series, as Google Trends sometimes yields ``empty'' results for entities with low search volume.
Time series were obtained with weekly granularity and span the entirety of 2019 and 2020.

\xhdr{Limitations} 
Our methodology for combining fact-checks with Google Trends data has some limitations.
First, since we map claims to one specific GKG entity and then query Google Trends, our methodology is unable to capture and cover complicated claims that may comprise multiple GKG entities
(recall that Google Trends does not support complicated queries that consist of multiple GKG entities). 
Also, a limitation that G-TAB does not address is that time series are normalized on a country-level basis (since, for each country, we build a different anchor bank).
Thus, to make cross-country comparisons, we additionally query for the entity ``Google'' (\texttt{/m/045c7b}), which is the world's most popular website. This allows us to consider the search volume of each query normalized by the attention going towards the ``Google'' entity (which is among the most popular in all countries).

\subsection{Misinformation-Induced Attention}
Ideally, to study fact-checking efforts leveraging Google Trends, we would like to obtain a metric that: 
\begin{enumerate}
\item  Measures the attention an entity (\eg, ``5G,'' \texttt{/m/0c41j8y/}) receives due to a piece of misinformation (\eg, 5G causes COVID\hyp{}19) and not due to other reasons (\eg, actual interest in the 5G technology). 
\item Is monotonically increasing over the relevant period so that we can meaningfully study \textit{when}, relative to a claim's attention cycle, it was fact-checked.
\item Is comparable across countries, even though G-TAB scales are country-specific.
\end{enumerate}

To obtain such a metric, we adjust G-TAB time series for the year 2020 by:
1)~subtracting a baseline calculated with data from the previous year; 
2)~enforcing the monotonicity of the metric by using the ramp (or ReLu) function; and
3)~dividing the values in the time series by a reference query that is popular across all countries (we use the search volume associated with the entity ``Google,'' \texttt{/m/045c7b}), as explained in \Secref{sec:fcgt}.

More specifically,
for each $\langle$entity,country$\rangle$ pair, in a given week $k$, let $v_k$ be the search volume estimated by Google Trends in that week in 2020, and $b$ be the average volume of the query in 2019.
We calculate the \emph{attention} up to given week $i$ as the sum of the positive differences between these two values up to week $i$,
normalized by $r$, the average search volume of the ``Google'' entity in 2020:
$$
 \text{Attention}(i) = \frac{1}{r}\sum_{k=1}^i \max(v_k - b, 0).
$$

We illustrate our metric in \Figref{fig:measures}. 
In (a), we show the Google Trends time series obtained with G-TAB for Brazil and Canada (solid lines) considering the entity `5G.' 
In 2020, we also show the baseline $b$, calculated as the average value of the time series in 2019 (dashed lines). The distance between these two lines (colored in the plot) is the quantity we sum in our attention metric ($v_k - b$).
In (b), we show the attention metric for the year 2020.
Notice that, in (a), the colored area (corresponding to the attention before normalization) is larger for Brazil than for Canada. However, this is not informative since the scales are country-specific. 
When we normalize the time series by the reference query (``Google''), we find that the increase in search volume was more expressive for Canada.
Normalizing also allows us to express the attention relative to the reference query. For example, in \textit{(b)} we find that, by the end of 2020, the attention 5G received in Canada was around 0.35. 
This means that the increased search interest received by the entity is equal to 35\% of the search volume of the reference query (``Google'') on an average day in 2020.

\xhdr{Total attention}
Note that we define our attention metric for each week of 2020. 
However, for our analyses, we consider the total attention received in 2020, \ie, the attention value for the last week of 2020 (since the metric is cumulative).
We refer to this value simply as \emph{total attention}. 
In \Figref{fig:measures}\textit{b}, this corresponds to the rightmost value of the time series for each country.

\xhdr{Relative attention} Another value of interest is the attention a claim receives up to a given moment in time relative to the total attention. For a given week $i$, we can obtain this value by simply dividing the attention at this week by the total attention:
$$
\text{Relative attention}(i) = \text{Attention}(i)/\text{Total attention}.
$$
This quantity is useful to assess \textit{when} claims were fact-checked. 
For example, in \Figref{fig:measures}\textit{b} we can calculate the relative attention to measure the fraction of the total attention the entity `5G' received in Brazil before it was first fact-checked (which happened on May 8, 2020). Doing so reveals that, by that time, the claim had already received 21\% of the total attention it would obtain throughout 2020.

\xhdr{Limitations}
A potential limitation of this methodology is that our baseline may not be a good counterfactual estimate of the volume in 2020 (\ie, the search volume had false information about the entities never existed).
However, employing a more sophisticated time series forecasting methodology (\eg, univariate unobserved components time series model~\cite{watson1986univariate}) yielded small differences, which led us to opt for the simpler option.
Another issue is that the reference query we use will invariably differ in popularity across different countries (\eg, searches related to Google may be more prevalent in Canada than in Brazil). We keep this limitation in mind as we devise our analyses in the forthcoming section.

\section{Results}

\begin{figure}[t]
\centering
\includegraphics[width=0.925\linewidth]{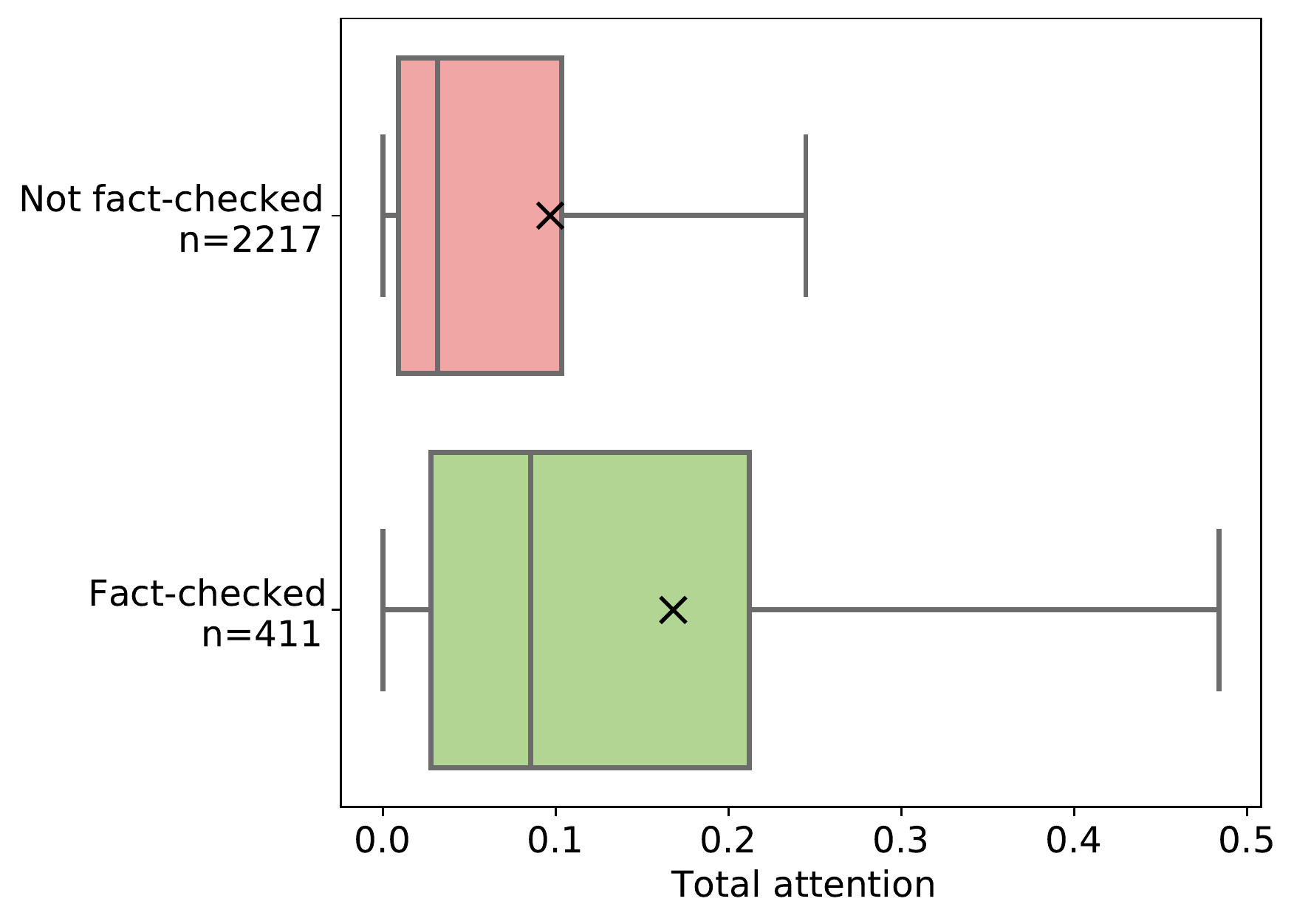}
\caption{Using a box plot, we depict the distributions of total attention for fact-checked \vs non-fact-checked claims. 
Box boundaries mark quartiles, the middle bar the median, and whiskers the 5th and 95th percentiles. The $x$ shows the mean of the distribution. Aggregation over countries was done by normalizing results by the same entity (as explained in \Secref{sec:methods}).}
\label{fig:relevance_1}
\end{figure} 

\subsection{Measuring the relevance of fact-checking}

We begin by characterizing \textit{what} was fact-checked as measured by Google Trends and fact-checking data from IFCN.

\xhdr{Fact-checked \vs non-fact-checked claims} 
Recall that each of the 39 claims we considered was fact-checked by a different set of countries. 
Here, we perform a series of analyses contrasting fact-checked \vs non-fact-checked claims.
In \Figref{fig:relevance_1} we show the distribution of total attention for fact-checked and non-fact-checked claims. 
Overall, we find that fact-checked claims received higher total attention ($\mu=0.168$; $SE=0.013$; $n=411$) than non-fact-checked claims ($\mu=0.097$; $SE=0.004$; $n=2217$). 

\begin{figure}[t]
\centering
\includegraphics[width=0.9\linewidth]{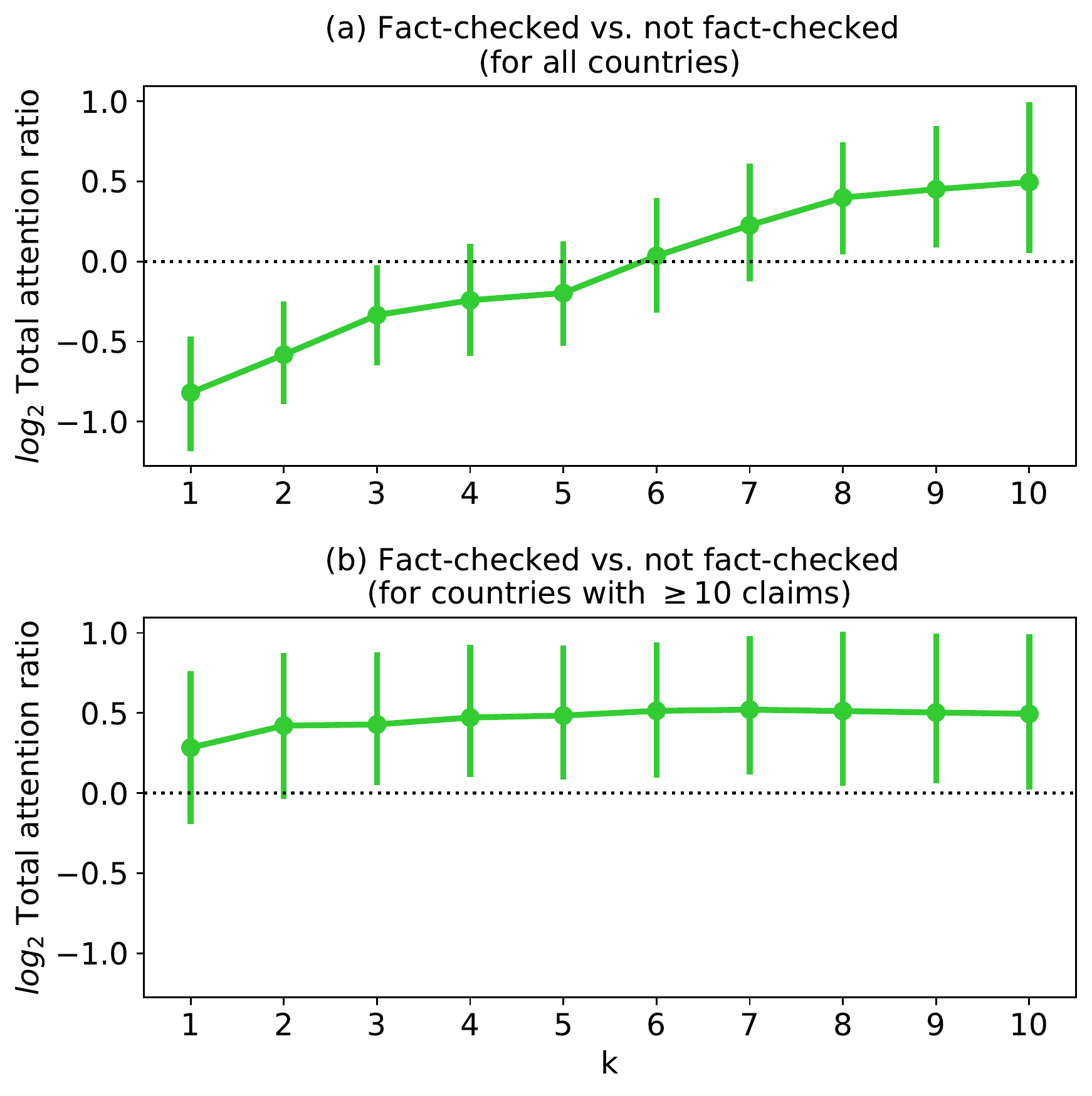}
\caption{
For each country, we compare the log-ratio between the total attention of the top $k$ fact-checked \vs  not fact-checked claims.
We depict the average of such values ($y$-axis) for varying values of $k$ ($x$-axis).
We consider two scenarios: in \textit{(a)} we study all countries, and the sample size shrinks as $k$ increases; in \textit{(b)} we consider countries with 10 or more fact-checked claims. Error bars represent 95\% CIs.
}
\label{fig:relevance_1a}
\end{figure}

In \Figref{fig:relevance_1a} we conduct a more nuanced analysis. 
For each country, we compare the total attention of the top $k$ fact-checked claims and not fact-checked claims that received the most attention. 
We sum the total attention going towards each of these claims and then calculate the log-ratio:
\begin{multline*}
\log_2 \frac{\sum_{i=1}^k \text{Total Att. of top i-th fact-checked claim}}
{\sum_{i=1}^k \text{Total Att. of top i-th non-fact-checked claim}}
\end{multline*}

Note that this ratio provides an intuitive interpretation.
If the total attention received by fact-checked and non-fact-checked claims is the same, the log-ratio will be 0;
If the total attention received by fact-checked claims is twice the total attention received by non-fact-checked claims, it will be 1; and if half as big, -1.

We conduct an analysis of the previously defined ratio in two different scenarios. First, in \Figref{fig:relevance_1a}\textit{a} we consider all countries, second, in  \Figref{fig:relevance_1a}\textit{b} we consider only countries that had at least 10 claims fact-checked. 
Note that in scenario (a), the number of samples being analyzed decreases as $k$ increases since we do not consider countries with less than $k$ fact-checks. 
However, in scenario (b), the sample size remains exactly the same for all values of $k$ ($n=13$). 

Considering scenario (a), we find that, for the very top claims ($k \in 1..5$), the previous finding does not hold: the fact-checked claims receive significantly \emph{less} total attention than the non-fact-checked claims.
However, as $k$ grows larger ($k \in 6..10$), this result is reversed and fact-checked claims receive significantly \emph{more} total attention (note that the averages are above the dashed line at $y=0$).
For $k=1$, for instance, we find that the top fact-checked claim was, on average, 44.4\% less popular than the non-fact-checked claim (as the log-ratio is -0.81 and $2^{-0.81}= 0.566$).

Scenario (b) sheds light on why this happens.
For countries where many claims are fact-checked ($\geq$ 10 claims fact-checked), fact-checked claims received \emph{more} total attention than non-fact-checked claims. 
This suggests that fact-checks in countries where fewer claims are fact-checked often address claims that are not receiving substantial attention.
Nevertheless, even for countries where more than 10 claims were fact-checked, many claims that received substantial attention were not.
In \Tabref{tab:countries_fc} we show, for these countries, what percentage of the 10 claims that received the most total attention were fact-checked.
For many countries (5/13, roughly 40\%), this percentage was not higher than 50\%, suggesting that relevant claims may have escaped the radar of fact-checking organizations.

\begin{table}[t]
\centering
\small

\caption{We show the percentage of the fact-checked claims among the top 10 claims with the largest total attention. We consider only countries that fact-checked at least ten claims.}
\begin{tabular}{llrl}
\toprule
Code &        Country &  \% of top 10 claims fact-checked  \\
\midrule
AR &      Argentina &        50\% &   \\
BR &         Brazil &        80\% &   \\
CO &       Colombia &        60\% &   \\
ES &          Spain &        70\% &   \\
FR &         France &        60\% &   \\
IN &          India &        70\% &   \\
IT &          Italy &        50\% &   \\
KE &          Kenya &        40\% &   \\
MX &         Mexico &        70\% &  \\
PH &    Philippines &        60\% &   \\
PT &       Portugal &        50\% &   \\
TR &         Turkey &        50\% &   \\
US &  United States &        70\% &   \\
\bottomrule
\end{tabular}
\label{tab:countries_fc}
\end{table}
\begin{figure}[t]
\centering
\includegraphics[width=\linewidth]{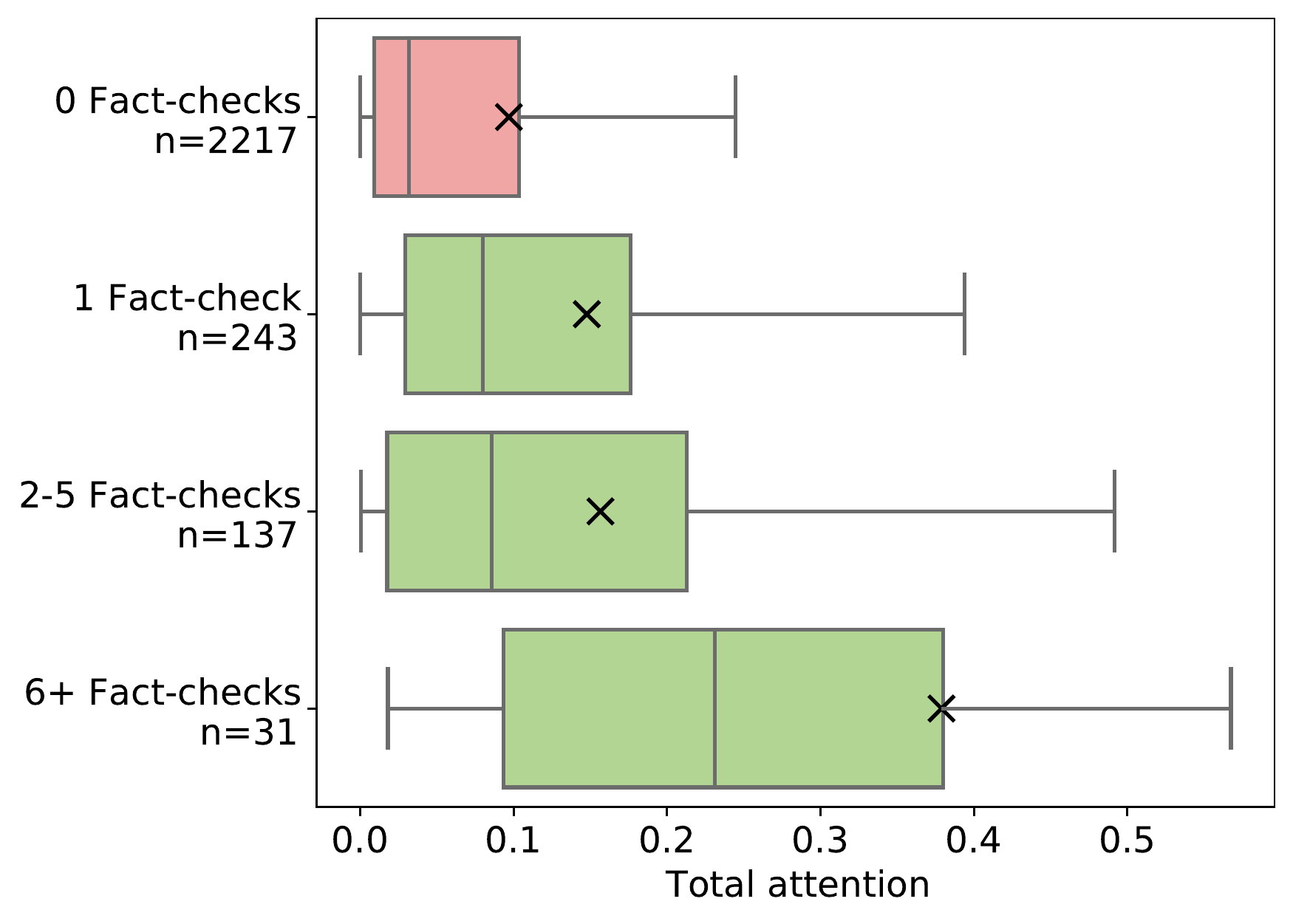}
\caption{We depict the distributions of total attention claims that received different number of fact-checks. 
Box boundaries mark quartiles, the middle bar the median, and whiskers the 5th and 95th percentiles. The $x$'s show the means of each distribution.}
\label{fig:relevance_2}
\end{figure}

\xhdr{Multiple fact-checking} 
In the previous analyses, we contrasted fact-checked and not fact-checked claims.
However, among fact-checked claims, some are fact-checked many more times than others. 
For example, in our dataset, there were 23 fact-checks debunking conspiracy theories associating COVID\hyp{}19 with the 5G technology in the United States. The claim that aspirin cures COVID\hyp{}19, however, was only fact-checked once (in the United States for the IFCN database we considered).

In Figure~\ref{fig:relevance_2} we further explore the relationship between the number of times a claim was  fact-checked and the total attention it received. 
We show the distribution of claims that were not fact-checked along with the distributions for claims that received 1, 2 to 5, and 6+ fact-checks.
Overall, we again observe that claims that were fact-checked received significantly more total attention than those that were not.
We find little difference in the mean value of received total attention received by claims that were fact-checked only once ($\mu=0.147$; $SE=0.014$; $n=243$) and those that were fact-checked between 2 and 5 times ($\mu=0.157$; $SE=0.019$; $n=137$).
However, claims that were fact-checked more than 6 times received substantially more total attention ($\mu=0.38$; $SE=0.10$; $n=31$).

To gain further insight into the relationship between multiple fact-checking and total attention, we calculate, for each country, the log-ratio between the attention received by the top $k$ most fact-checked claims and the top $k$ fact-checked claims that received the most attention.
Notice that here, unlike in \Figref{fig:relevance_1a}, we are comparing two sets of fact-checked claims.
If the set of top $k$ claims that were the most fact-checked for a given country is the same as the set of top $k$ claims that received the most attention, the log-ratio equals 0. 
However, it may be that there exists claims that are in the top $k$ for total attention but not in the top $k$ for most fact-checked.  
In such cases, we will observe a log-ratio smaller than 0.

\begin{figure}[t]
\centering
\includegraphics[width=\linewidth]{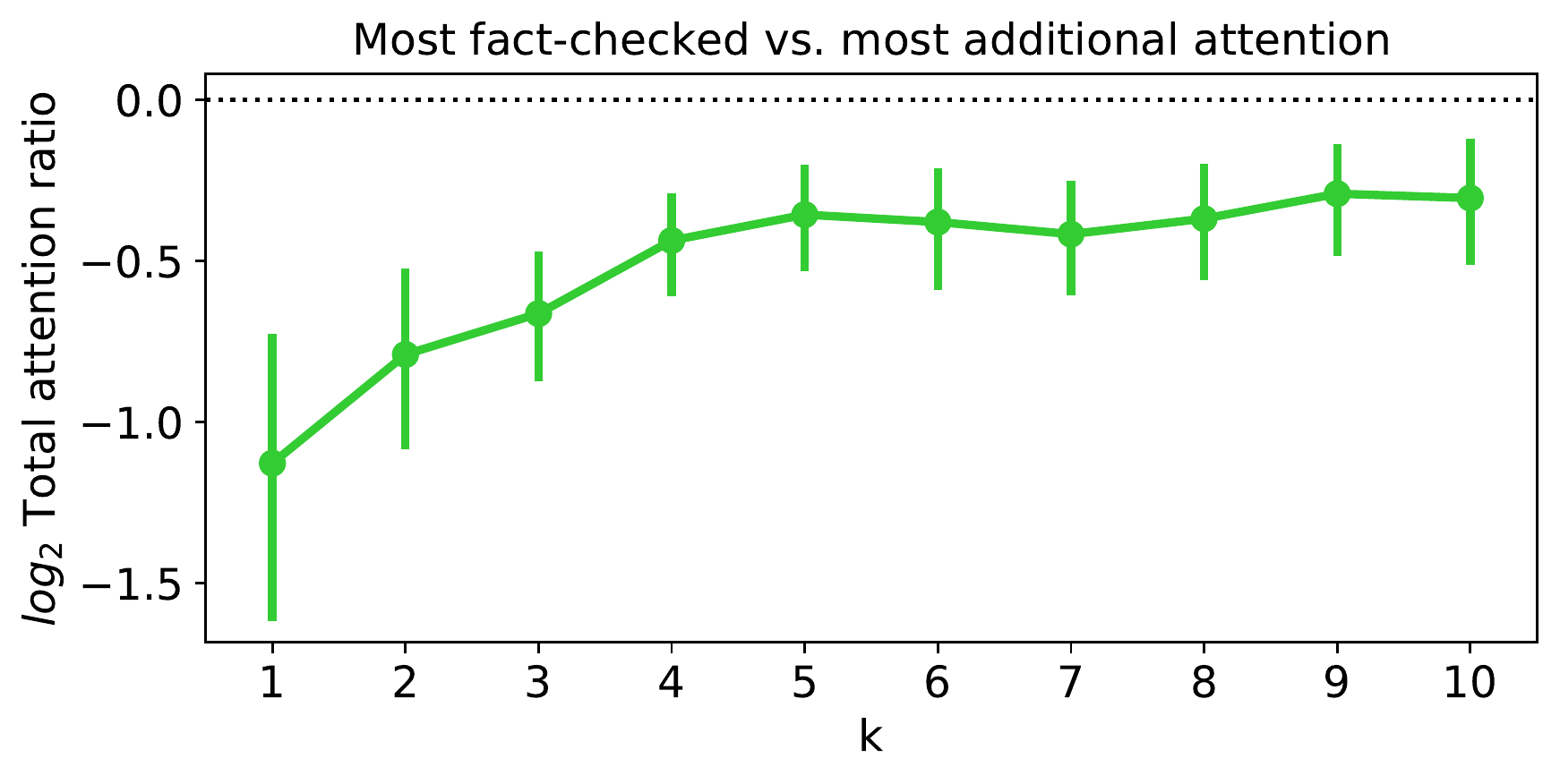}
\caption{For each country, we compare the log-ratio between the total attention of the $k$ most fact-checked claims and the $k$ fact-checked claims that received the largest total attention.
We depict the average over countries of such values ($y$-axis) for varying values of $k$ ($x$-axis). Error bars represent 95\% CIs.}
\label{fig:relevance_2a}
\end{figure}

In \Figref{fig:relevance_2a} we show the average value for the aforementioned ratio for $k$ between 1 and 10. 
For $k=1$, we find that the most fact-checked claim receives on average, half of the total attention received by the most popular claim (as the log-ratio is $-1.12$, and $2^{-1.12}= 0.46$). 
As $k$ grows larger, the average log-ratio approaches $-0.4$, suggesting that the most fact-checked claims have around 75\% of the attention of the claims that received the most attention. 
Here, we find little difference when analyzing only countries with more than 10 fact-checked claims.

\begin{figure}[t]
\centering
\includegraphics[width=0.925\linewidth]{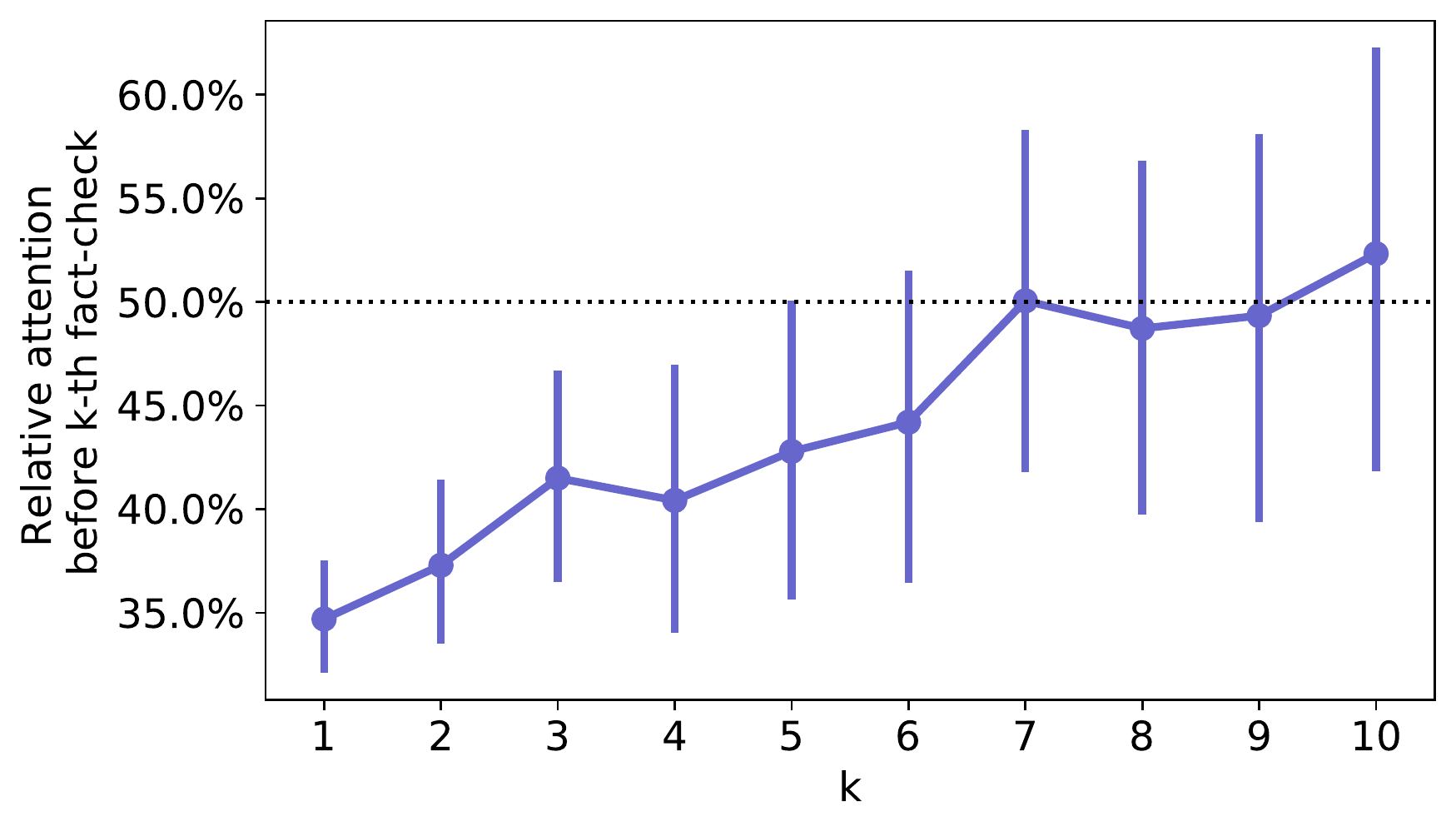}
\caption{We depict the mean relative attention ($y$-axis) received by fact-checked claims when they received their $k$-th fact-check ($x$-axis). We vary $k$ between 1 and 10. Error bars indicate 95\% CIs.}
\label{fig:speed_1}
\end{figure}

\subsection{Measuring the speed of fact-checking}

We now turn to characterize \textit{when} in the life cycle of a claim it gets fact-checked. 

\xhdrNoPeriod{When are claims fact-checked?} 
In \Figref{fig:speed_1}  we examine when claims were fact-checked. 
For all fact-checked claims, we depict the relative attention received when the claim was fact-checked for the $k$-th time in a given country. 
We find that the first fact-check happens, on average, when the claim has received 35\% ($\mu=0.346$; $SE=0.012$; $n=410$) relative attention, \ie 35\% of the total attention it would eventually receive over the year of 2020. 
As we increase $k$, this value grows up to when $k=7$, when it reaches around 50\% ($\mu=0.501$; $SE=0.041$; $n=28$). 
For $k\in8..10$, it continues around this value, marked in the plot by a dotted line. Only 11 $\langle$claim,country$\rangle$ pairs were fact-checked 10  or more times.

\begin{figure*}[t]
\begin{minipage}{.5\textwidth}
\centering
\includegraphics[width=\linewidth]{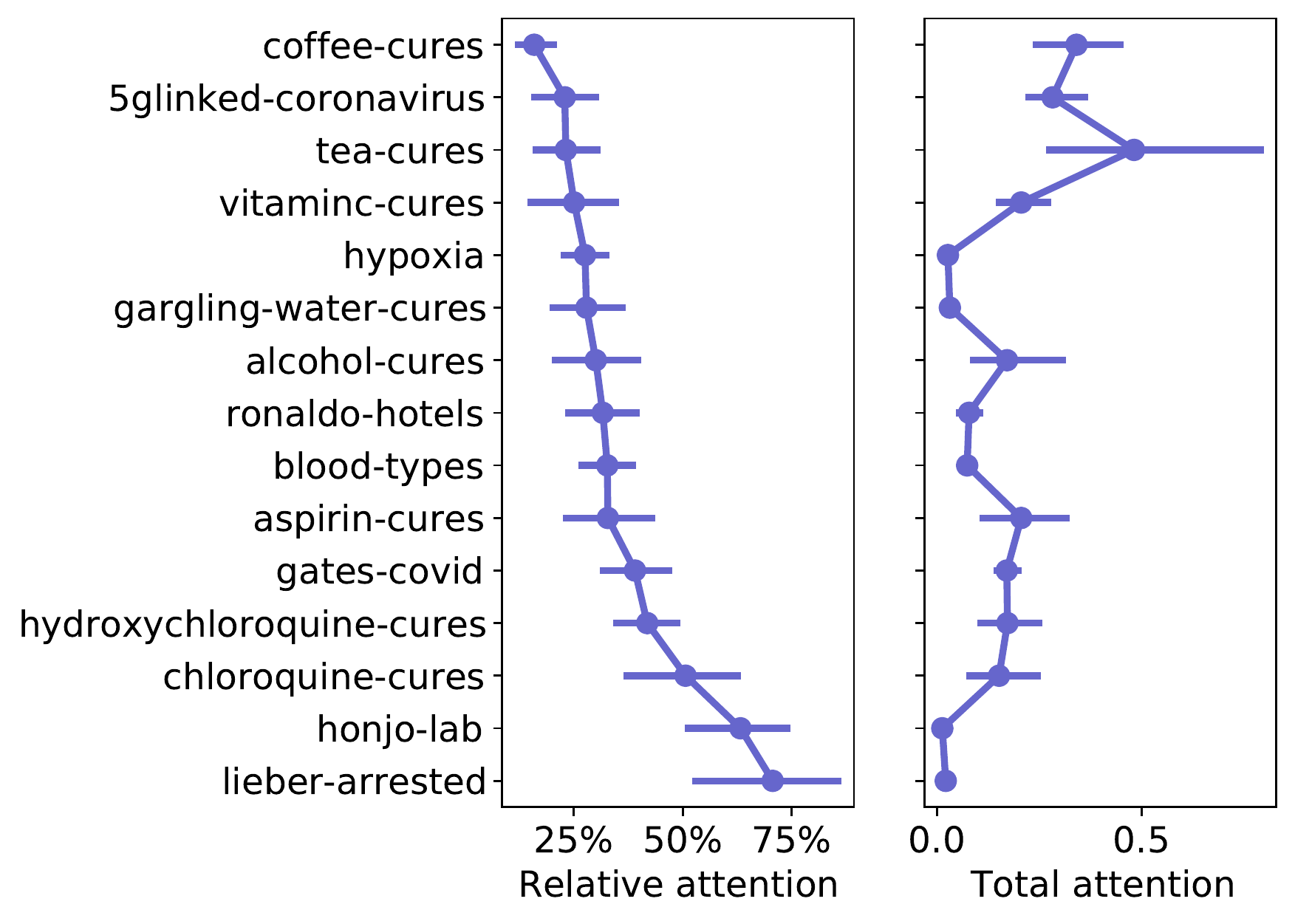}
\end{minipage}%
\begin{minipage}{0.5\textwidth}
\centering
\includegraphics[width=\linewidth]{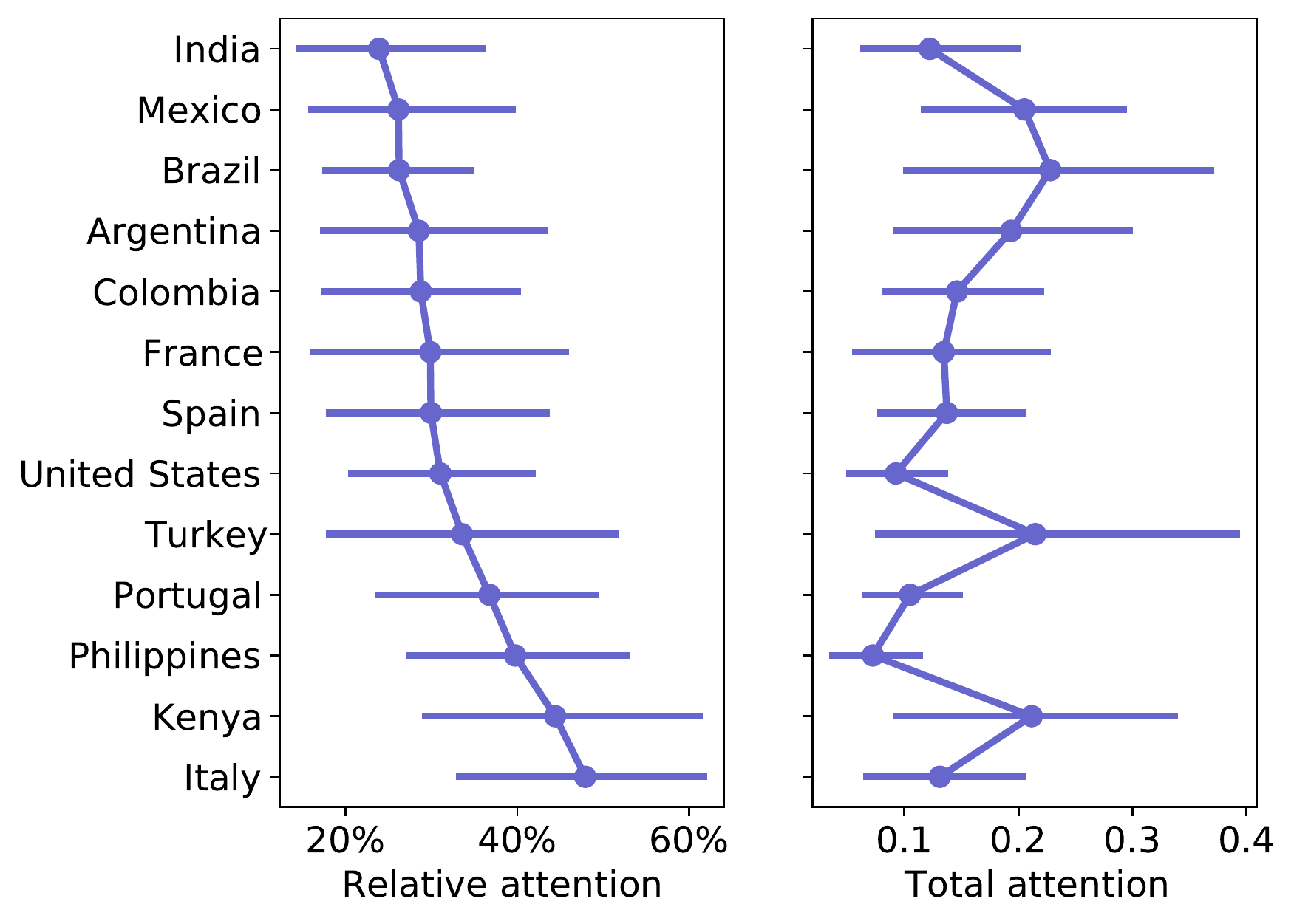}
\end{minipage}
\caption{\textit{(a)} For claims that were fact-checked in at least 10 countries, we show mean values for the relative attention at the time of the first fact-check (left); and the total attention received by the claim in 2020 (right).
\textit{(b)} We conduct the same analysis considering the countries where  at least 10 claims were fact-checked. 
Error bars represent 95\% CIs.
}
\label{fig:speed_2}
\end{figure*}

\xhdrNoPeriod{When are different claims first fact-checked?}
We find that different claims are first  fact-checked at distinct moments of their life cycle. 
In \Figref{fig:speed_2}\textit{a} we show, for claims that were fact-checked in more than 10 countries, the average relative attention at the time of the first fact-check (on the left) and the average total attention these claims received (on the right).
While some claims were fact-checked very soon relative to the total attention received (\eg, coffee cures coronavirus was first fact-checked with 15.9\% relative attention), others were fact-checked after receiving most of the attention they would receive in 2020 (\eg, the claim that Charles Lieber, Harvard professor, had been arrested for creating COVID\hyp{}19 was fact-checked with 70.6\% relative attention).

Relative attention at the time of the first fact-check is negatively correlated with the total attention these claims received, as can be seen on the right-hand side of \Figref{fig:speed_2}\textit{a} (Spearman's $\rho=-0.62$; $p=0.013$). 
Additionally, we find that there is no significant correlation between the date in 2020 when the claim was first fact-checked and the relative attention ($\rho=0.18$; $p=0.52$).
An Analysis of Covariance (ANCOVA) considering the relative attention as the response variable, claims as categorical independent variables, and the total attention received in 2020 and day of first fact-checked as a covariate indicates that there is a significant difference among when claims were fact-checked ($F=6.46$; $p<0.001$).

\xhdrNoPeriod{When do different countries first fact-check?}
We find no significant difference in when countries first fact-check claims.
In \Figref{fig:speed_2}\textit{b} we show, for countries that fact-checked at least 10 claims, the average relative attention at the moment they first fact-checked a claim (on the left) and the average total attention received at the time of the first fact-check (on the right).
We find that differences in when countries fact-check are much more subtle, as can be noticed by the wider and overlapping confidence intervals in the left-hand side of \Figref{fig:speed_2}\textit{b}. 
Also, the correlation between total attention and relative attention is weaker and non-significant ($\rho=-0.42$; $p=0.16$).
To determine whether there are distinct differences between the averages of each country, we again run our Analysis of Covariance now considering the country as the  categorical independent variable.
Here, we find no significant difference between the average values of each country ($F=1.09$; $p=0.37$).

\section{Discussion}

We propose a quantitative framework to characterize fact-checking efforts using online attention and apply it to a dataset of COVID\hyp{}19 related fact-checking traces from organizations affiliated with the International Fact-checking Network~\cite{ifcn}.
Although we focus on Google Trends, we argue that our framework could be easily extended to other online attention signals (\eg, Wikipedia pageviews).

Our analysis revealed a disconnect between online attention and fact-checking efforts.
According to our framework, many claims receive substantial online attention but are not fact-checked. 
The attention going towards non-fact-checked claims is particularly high in countries where fewer claims were fact-checked.
Evidently, the attention received by a claim is only one of many factors that fact-checking organizations take into account. Organizations may also prioritize checking claims that are more likely to be harmful and practical to check~\cite{fbfc}. 
Nevertheless, previous work has investigated motivations behind fact-checking~\cite{graves2016understanding}, finding that it is strongly driven by professional motives within journalism rather than by audience demand. 
Given this reality, we argue that news organizations may benefit by incorporating online attention signals to their fact-checking pipeline.

The proposed framework also allows the characterization of \textit{when} claims are fact-checked. 
We find that most  fact-checks are issued when claims had already received, on average, 35\% of the total attention they would eventually receive in 2020. This value, however, varies widely by claim, suggesting that fact-checking organizations do not systematically check claims at a specific moment of their life cycle.
Further, recent work suggests that timing matters when correcting misconceptions~\cite{brashier_timing_2021}, and, in this context, online attention could be used to efficiently promote fact-checks.
Note that this would require additional information on how fact-checks are leveraged by platforms and news organizations, which was not available in the data at hand.

The usage of online attention may also be beneficial to cross-country collaborative fact-checking efforts.
Misinformation respects no national borders, and, as such,  rumors will inevitably migrate across countries and languages.
Organizations such as the International fact-checking Network aim, among other things, to reduce the burden of fact-checking the same rumor multiple times across different countries. Within this context, online attention may play a variety of roles. 
First, it can be used as an ``early warning'' system. 
For example, suppose a fact-checking organization in a given country attributes entities to their fact-checks. 
Then, other organizations could be quickly informed if the entity's attention increases in their respective countries.
Second, it can be used to trace global misinformation pathways. Analyzing the attention time series obtained, one could investigate from which countries misinformation migrates from and towards.

Although Google Trends is a ``battle-tested'' metric for online attention, there are still limitations associated with its usage. 
First, the popularity of Google varies across countries, and the attention signal we are studying is bound to be an imperfect proxy for all attention a claim receives (\eg, on other platforms such as WhatsApp or Facebook). 
Future work could further examine the relationship between Google Trends and the ``real'' attention a claim receives, which could be measured, for instance, using surveys. 
Adjusting for confounding factors such as the level of Internet literacy and the popularity of Google in different countries may help to obtain better estimates of the level of exposure specific claims receive across the globe.
Second, in the proposed framework, we consider attention beyond a baseline to be associated with misinformation. 
This is an inexact approximation, and, given a specific entity, a boost in online attention may be due to non-misinformation-related reasons. 
In this context, future work could try to build better counterfactual estimates of how much attention a claim would have received in the absence of misinformation.

Overall, we argue that incorporating online attention in the study and the practice of fact-checking is an important step towards the systematization of fact-checking efforts and towards confirming the ecological validity of findings obtained in controlled, and to some extent, non-realistic scenarios~\cite{nyhan_why_2021}.


{
\small
\bibliography{refs}
 }

\end{document}